\documentclass[twocolumn]{aastex631}
\usepackage{chngcntr}
\usepackage{booktabs}
\usepackage{xspace}
\usepackage{gensymb}
\usepackage{amsmath}
\usepackage{orcidlink}

\newcommand{\seff}{$S_{\text{eff}}$} 
\newcommand{\teff}{$T_{\text{eff}}$} 
\newcommand{\pir}{\textit{P-I} }

\newcommand{\rrate}{$R_{\mathrm{rate}}\,\, $}
\newcommand{\pmax}{$P_{\mathrm{rate}}^{\mathrm{max}}\,\,$}

\newcommand{\photounits}{$\mu\mathrm{mol\,\, O}_2 \,\, \mathrm{mg}^{-1} \,\, \mathrm{h}^{-1}\,\, $}

\counterwithout{figure}{section}

\graphicspath{ {./figures/} }
\submitjournal{ApJL}

\shorttitle{Photosynthetic Habitable Zone}
\shortauthors{Hall et al.}

\begin{document}

\title{A New Definition of Exoplanet Habitability: Introducing the Photosynthetic Habitable Zone}

\correspondingauthor{C. Hall}
\email{ch27976@uga.edu}

\author[0000-0002-8138-0425]{C. Hall}
\affil{Department of Physics and Astronomy, The University of Georgia, Athens, GA 30602, USA.} 
\affil{Center for Simulational Physics, The University of Georgia, Athens, GA 30602, USA.}

\author[0000-0003-4661-6735]{P. C. Stancil}
\affil{Department of Physics and Astronomy, The University of Georgia, Athens, GA 30602, USA.} 
\affil{Center for Simulational Physics, The University of Georgia, Athens, GA 30602, USA.}

\author[0000-0002-8590-7271]{J. P. Terry}
\affil{Department of Physics and Astronomy, The University of Georgia, Athens, GA 30602, USA.}
\affil{Center for Simulational Physics, The University of Georgia, Athens, GA 30602, USA.}

\author[0000-0003-3259-079X]{C. K. Ellison}
\affil{Department of Microbiology, The University of Georgia, Athens, GA 30602, USA.}


\begin{abstract}
It may be possible to detect biosignatures of photosynthesis in an exoplanet's atmosphere. However, such a detection would likely require a dedicated study, occupying a large amount of telescope time. It is therefore prudent, while searching for signs of life that we may recognise, to pick the best target possible. In this work, we present a new region,  the ``photosynthetic habitable zone'' \textemdash the distance from a star where both liquid water and oxygenic photosynthesis can occur. It is therefore the region where detectable biosignatures of oxygenic photosynthesis are most likely to occur.
%
%
Our analysis indicates that in the most ideal conditions for life and no atmospheric effects, the photosynthetic habitable zone is almost as broad as the habitable zone. On the other hand, if conditions for life are anything less than excellent and atmospheric effects are even moderate, the photosynthetic habitable zone is concentrated at larger separations around more massive stars. Such cases are also not tidally locked to their host star, which could result in planetary rotation periods similar to the Earth's. We identify five planets, Kepler-452 b, Kepler-1638 b, Kepler-1544 b and Kepler-62 e and Kepler-62 f, that  are consistently in the photosynthetic habitable zone for a variety of conditions, and we predict their day lengths to be between 9 and 11 hours. We conclude that the parameter space in which we should search for signs of life is much narrower than the standard habitable zone.

\end{abstract}

\section{Introduction}

\noindent Since the first exoplanet atmospheric detection around a $1.35$ R$_\mathrm{J}$ planet \citep{charbonneau2002}, astronomers have been pushing the limits to smaller and smaller planets, such as the detection of water vapour around the $8 \mathrm{M}_\mathrm{E}$ planet K12-18b \citep{8me}. The continued discovery of exoplanet atmospheres, and increasing technological capability, has raised the prospect of finding a planet that may be inhabited by life. Subsequently, it has been determined that the characterization and detection of biosignatures \textemdash atmospheric spectral features that could indicate signs of life on a planet \textemdash should be an area of focus for astrobiology \citep[see, e.g.][]{seager2012,kaltenegger2017,lammer2019}.

 To date, over 5000\footnote{https://exoplanetarchive.ipac.caltech.edu/} exoplanets have been discovered using a mix of ground-based and space-based methods. With the successful launch of JWST and future observatories such as the European Extremely Large Telescope (E-ELT) and the Thirty Meter Telescope (TMT), we are moving from the era of exoplanet discovery to exoplanet atmospheric characterization. However, characterizing these worlds remains an enormous challenge. For example,  the most promising O$_2$ feature for JWST appears to be the O$_2\rightarrow X$ collisional induced adsorption band at $6.4$ $\mu$m \citep{fauchez2020}, but even for a target such as TRAPPIST 1-e \citep{gillon2016temperate,gillon2017seven} this would require more than 700 transits, longer than the anticipated lifetime of JWST given that TRAPPIST 1-e has a 6 day orbital period and is only visible to JWST for less than a third of the year \citep{gillon2020}. Even the more favourable strong O$_3$ band at 10$\mu$m would require more than 100 transit observations on a planet such as TRAPPIST 1-e \citep{lin2021} to detect it at just 3$\sigma$.

Fortunately, O$_2$ and O$_3$ are not the only biosignatures. More generally, atmospheric chemical disequilibrium, characterised by the coexistence of two or more long-term incompatible gases \citep{lovelock1965,sagan1993,cockell2009}, can be considered a sign of ongoing life. The Archean Earth had a biogenic disequilibrium caused by the coexistence of N$_2$, CH$_4$, CO$_2$, and liquid water, which could be possible to remotely detect on an Earth-sized planet \citep{disequilibrium}. Simultaneous detection of abundant CH$_4$ and CO$_\mathrm{2}$ is therefore considered a biosignature. Happily, detecting this CH$_4$-CO$_2$ pair is feasible, requiring $\sim5-30$ co-added JWST transits depending on if the stratosphere is dry or has a cloud or haze layer \citep{mikal2022}.

Observing resources are valuable and finite, so choosing the best targets to search for biosignatures of any kind is imperative. The main criterium is whether the planet can sustain liquid water on its surface by residing an appropriate semi-major axis from its host star, referred to as the habitable zone \citep{oghz,kasting1993}. However, liquid water alone is not enough for life.  Life requires energy to remain out of equilibrium with its environment. For almost all the biomass on Earth, this energy source is oxygenic photosynthesis \citep{bar-on2018}. 
We therefore suggest that a new criterium be used to determine where biosignatures may be found. In this work, we demonstrate that, like the habitable zone, the \textit{photosynthetic habitable zone} is a bounded strip on a plot of stellar mass against semi-major axis. It occurs where both liquid water and photosynthesis is simultaneously possible. It is where the search for life in the Universe should be concentrated under the assumption of biosignatures similar to those generated by past or present Earth. 

We detail our calculations of the habitable zone in Section \ref{subsec:hz}, our calculations of photosynthesis rate curves in Section \ref{subsec:ratecurves}, and the photosynthetic habitable zone in Section \ref{subsec:bz}. We present our results in Section \ref{sec:results} and discuss assumptions and limitations in Section \ref{sec:discussion}. We summarise and present our conclusion in Section \ref{sec:conclusion}.

\section{Methods}
\label{sec:methods}

\subsection{The Habitable Zone}
\label{subsec:hz}
\noindent We used pre-main sequence (PMS) evolutionary models \citep{baraffe2015}\footnote{\url{http://perso.ens-lyon.fr/isabelle.baraffe/BHAC15dir/BHAC15_iso.2mass}} to obtain stellar effective temperature, \teff,  as a function of stellar mass, $M_*$. We assumed an age of 1 Gyr, corresponding to the approximate time primitive life first appeared on Earth \citep[e.g.][]{2017Natur.543...60D,Cavalazzi2021}. The habitable zone (HZ) for each stellar mass was calculated using the method described in \citet{kopparapu2013}. We use their derived relationships between HZ  effective temperature,  $T_\mathrm{eff}$ and stellar fluxes, \seff,  in the range 2600 K $\leqslant T_{\text{eff}} \leqslant 7200  $K:
\begin{equation}
\label{eq:seff}
S_{\text {eff }}=S_{\text {eff } \odot}+a T_{\star}+b T_{\star}^{2}+c T_{\star}^{3}+d T_{\star}^{4},
\end{equation}
where $T_{\star}=T_{\text{eff}}-5780$ K,  and 
\begin{equation}
S_{\text {eff } \odot} = \frac{L_\odot}{4\pi R_\odot^2 \sigma T_\odot^4}, 
\end{equation}
where $L$ is stellar luminosity and $R$ is stellar radius. The coefficients $a,b,c$ and $d$ are determined by scenario, for example runaway greenhouse (Inner HZ) and maximum greenhouse (outer HZ). We used updated coefficient values published online \footnote{{\url{http://depts.washington.edu/naivpl/sites/default/files/hz_0.shtml\#overlay-context=content/hz-calculator}} } as per \citet{erratum}, and detail them in Table \ref{tab:coeff}. The corresponding HZ distance for a given star is then  
\begin{equation}
d=\left(\frac{L / L_{\odot}}{S_{\mathrm{eff}}}\right)^{0.5} \mathrm{AU},
\end{equation}
where $L / L_{\odot} $ is the luminosity of the star compared to the Sun. 


\begin{table}
    \centering
    \begin{tabular}{ccc}
    \hline
    \hline
        Constant & Inner HZ & Outer HZ \\ 
                        &(Runaway Greenhouse) & (Max. Grenhouse ) \\ \hline

        $S_{\text {eff } \odot}$ & 1.107 & 0.356 \\ 
        $a$ & $1.332 \times 10^{-4}$ & $6.171\times 10^{-5}$ \\ 
        $b$ & $1.580\times 10^{-8}$ & $1.698\times 10^{-9}$ \\
        $c$ & $-8.308\times 10^{-12}$ & $-3.198\times 10^{-12}$ \\ 
        $d$ & $-1.931\times 10^{-15}$ & $-5.575\times 10^{-16}$ \\ 
\hline
    \end{tabular}
    \caption{Coefficients used in Eq. \ref{eq:seff} to calculate habitable stellar fluxes, and corresponding habitable zones.\label{tab:coeff}}
\end{table}


\subsection{The Photosynthesis zone}
\label{subsec:ratecurves}
%
\noindent Photosynthesis is the chemical reaction by which organisms use energy from sunlight to synthesis sugar from carbon dioxiode and water:
\begin{equation}
6 \mathrm{CO}_{2}+6 \mathrm{H}_{2} \mathrm{O} \xrightarrow{h\nu}   \mathrm{C}_{6} \mathrm{H}_{12} \mathrm{O}_{6}+6 \mathrm{O}_{2}.
\end{equation}
Three variables directly impact the rate of photosynthesis \citep[][]{gaastra1959photosynthesis}: light intensity ($I$), temperature ($T$), and carbon dioxide (CO$_2$). Water availability indirectly impacts the rate of photosynthesis, as water stress causes plant structures to wilt, reducing CO$_2$  availability \citep{muller2011water}. In this work, we make two assumptions: 1) that the photosynthetic life we consider is ocean based, and so has an unlimited reservoir of water and 2) the global average of CO$_2$ concentration is sufficiently high that photosynthesis is not rate-limited by CO$_2$ concentration. Today's CO$_2$ levels are $\sim 0.04$\% by volume, and the early Archaen Earth had an estimated atmospheric CO$_2$ concentration of 70\% by volume \citep[see, e.g.,][]{doi:10.1126/sciadv.aay4644}, so this assumption is probably reasonable.


The chlorophyll $a$-normalised net photosynthetic rate, $P$, as a function of irradiance intensity, $I$, is given by \citep{model3}:

\begin{equation}
\label{eq:prate}
P(I)=\frac{I}{\alpha I^{2}+\beta I+\gamma}-R_{\mathrm{rate}}, 
\end{equation}
where $\alpha$ and $\beta$ are dimensionless parameters,  $\gamma$ is defined as the reciprocal of the light-limited initial slope of the \pir curve, and \rrate is the dark respiration rate, the minimum rate at which glucose is combined enzymatically with oxygen to release energy and CO$_2$. Net photosynthetic rate is the total output of molecular oxygen per unit biomass per unit time [\photounits]. The maximum photosynthetic rate is given by
\begin{equation}
\label{eq:pmax}
P_{\mathrm{rate}}^{\text{max}}=\frac{1}{\beta+2 \sqrt{\alpha \gamma}}-R_{\mathrm{rate}}.
\end{equation}
The parameters $\alpha,\beta$ and $\gamma$ in Equations \ref{eq:prate} and \ref{eq:pmax}  are determined by performing best-fit analysis to empirically determined photosynthesis rate curves of phytoplankton in the laboratory setting, the values used in this work are $\alpha=1.0\times 10^{-5}$, $\beta=1.0\times 10^{-3}$, and $\gamma=2.0$  \citep{yang2020}. Three values of the dark respiration rate are explored,  which determine the quality of conditions for life. As conditions for life become less favourable, \rrate becomes a greater fraction of $P^{\mathrm{max}}_\mathrm{rate}$ \citep{geider1989}. We consider excellent conditions to be Earth-like (\rrate = 0.3 \pmax), and optimistic and pessimistic conditions to be \rrate = 0.6 \pmax and \rrate = 0.8 \pmax respectively. 

\begin{figure}[h]
    \centering
    \includegraphics[width=\linewidth]{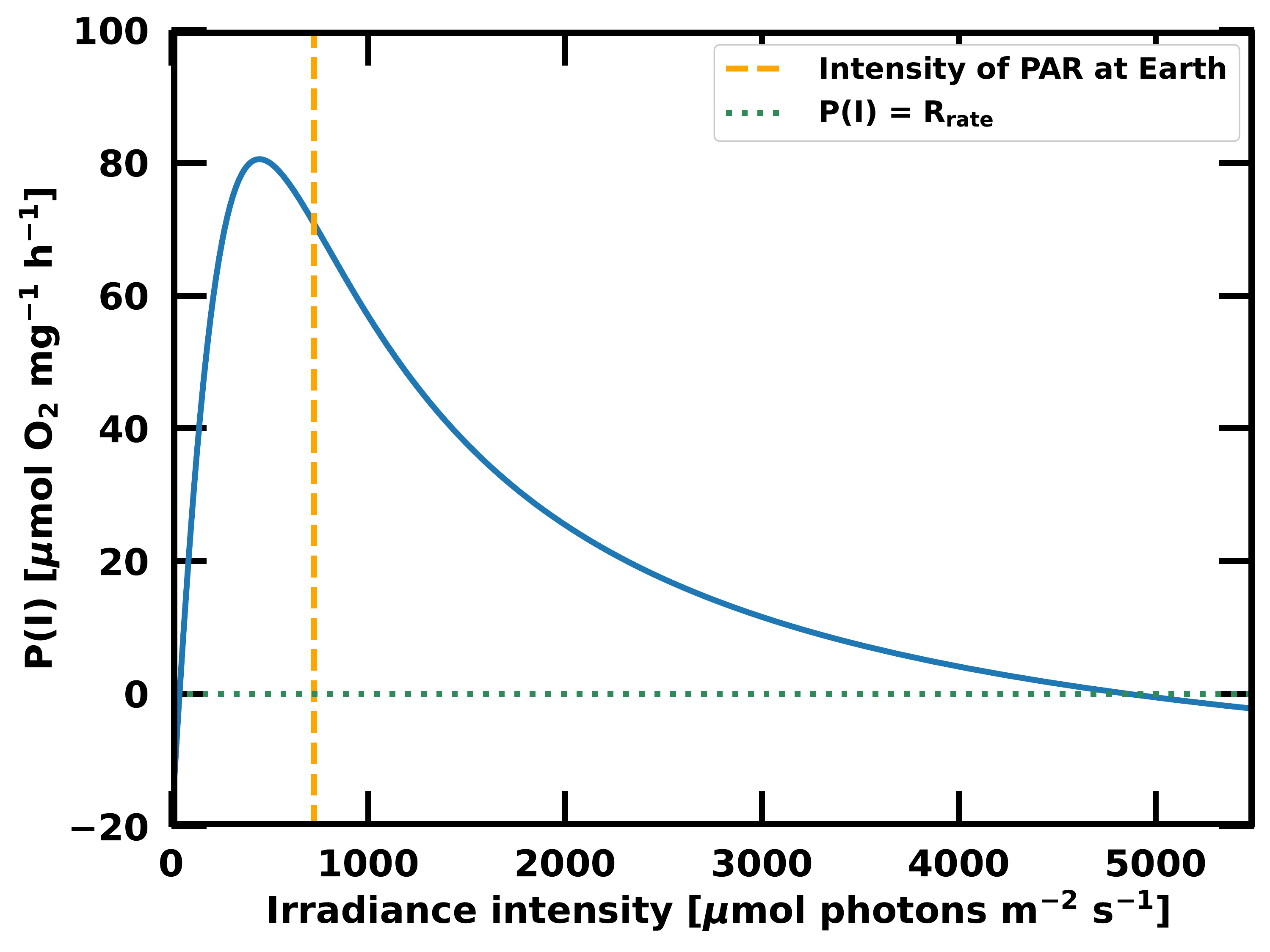}
    \caption{ Net photosynthetic rate versus irradiance intensity. Dotted green line indicates where rate of photosynthetic production of O$_2$ is equal to consumption of O$_2$ during respiration, such that net photosynthetic production of O$_2$ is zero. Rate of respiration is \rrate = 20 $\mu\mathrm{mol\,\, O}_2 \,\, \mathrm{mg}^{-1} \,\, \mathrm{h}^{-1}$. Orange vertical line is intensity of PAR at Earth.}
    \label{fig:photorate}
\end{figure}

The \pir curve from Equation \ref{eq:prate} is shown in Figure \ref{fig:photorate}. The green dotted line in Figure \ref{fig:photorate} shows the line where the net photosynthesis rate is $P(I) = 0 $, due to rate of primary production equalling the rate of respiration.  Any irradiance intensity that causes a total net negative photosynthesis rate will result in no net oxygen being produced. Therefore, for photosynthesis to increase atmospheric content of O$_2$, $P >0$ is the absolute lowest limit required. 

We term the region of parameter space where $P(I)>0$ (Equation \ref{eq:prate}) the ``photosynthesis zone''  (PZ) for convenience, and it is shown in pink in Figure. \ref{fig:bz}. Photosynthesis cannot proceed where $P\leq 0$, since the organism's respiration rate is exceeding that of its primary production.

\subsection{The Photosynthetic Habitable Zone}
\label{subsec:bz}



We suggest the existence of a \textit{photosynthetic habitable zone} (PHZ), a region of parameter space where the habitable zone overlaps the photosynthesis zone. It is in the photosynthetic habitable zone, rather than the habitable zone, that humanity should concentrate its search for spectral signs of life, since they can only be present where both liquid water and $P(I)>0$ occur. To obtain the region where $P(I)>0$, we calculate the irradiance intensity as a function of stellar mass and planet semi-major axis. We begin with the Planck function:
\begin{equation}
B(\lambda, T)=\frac{2 h c^{2}}{\lambda^{5}}   \bigg[e^{\frac{h c}{\lambda k_{\mathrm{B}} T}}-1\bigg]^{-1},
\end{equation}
and obtain the intensity of photosynthetically active radiation, $I_\mathrm{PAR}$, by integrating between $\lambda_\mathrm{min}=$400 nm and $\lambda_\mathrm{max}=$700 nm
\begin{equation}
\label{eq:I}
I_\mathrm{PAR}= \int_{\lambda_\mathrm{min}}^{\lambda_\mathrm{max}} \frac{2 h c^{2} / \lambda^{5}}{e^{h c / \lambda k_\mathrm{B} T}-1} d \lambda . 
\end{equation}
The number of photons emitted by the star per unit time, $\dot{N}_{\star}$, in this wavelength range is then obtained by multiplying Equation \ref{eq:I} by the surface area of the star, and dividing by the energy of each photon, so that we obtain 
\begin{equation}
\dot{N}_{\star}=4 \pi R_{\star}^{2} \int_{\lambda_{\min }}^{\lambda_{\max }} \frac{2 c}{\lambda^{4}}\left[\exp \left(\frac{h c}{\lambda k_{B} T_{\star}}\right)-1\right]^{-1} d \lambda.
\end{equation}
%
%
Assuming a circular orbit, the photon flux at the top of a planetary atmosphere, $\Phi$, at a distance $a$ from the star is
\begin{equation}
\label{eq:phi}
\Phi = \frac{\dot{N}_{\star}}{4 \pi a^{2}}.
\end{equation}
%
%
We account for atmospheric attenuation such that the light intensity at the bottom of the atmosphere is 
\begin{equation}
\label{eq:phi_attenuate}
I = f_\mathrm{a} \cdot \frac{\dot{N}_{\star}}{4 \pi a^{2}} 
\end{equation}
where $f_\mathrm{a} \leq 1.0$ is the fractional attenuation due to atmospheric effects. We consider three attenuation efficiencies: no attenuation, $f_\mathrm{a} = 1.0$, moderate attenuation $f_\mathrm{a} = 0.6$, and Earth-like attenuation, $f_\mathrm{a} = 0.2$ \citep{oceandynamics,lingamloeb2021}. The intensity in Equation \ref{eq:phi_attenuate} is used in Equation \ref{eq:prate} to obtain $P(I)$ that would be achieved by a phytoplankton-like species as a function of stellar mass and planet semi-major axis. 

\section{Results}
\label{sec:results}

\begin{figure*}
    \centering
    \includegraphics[width=\textwidth]{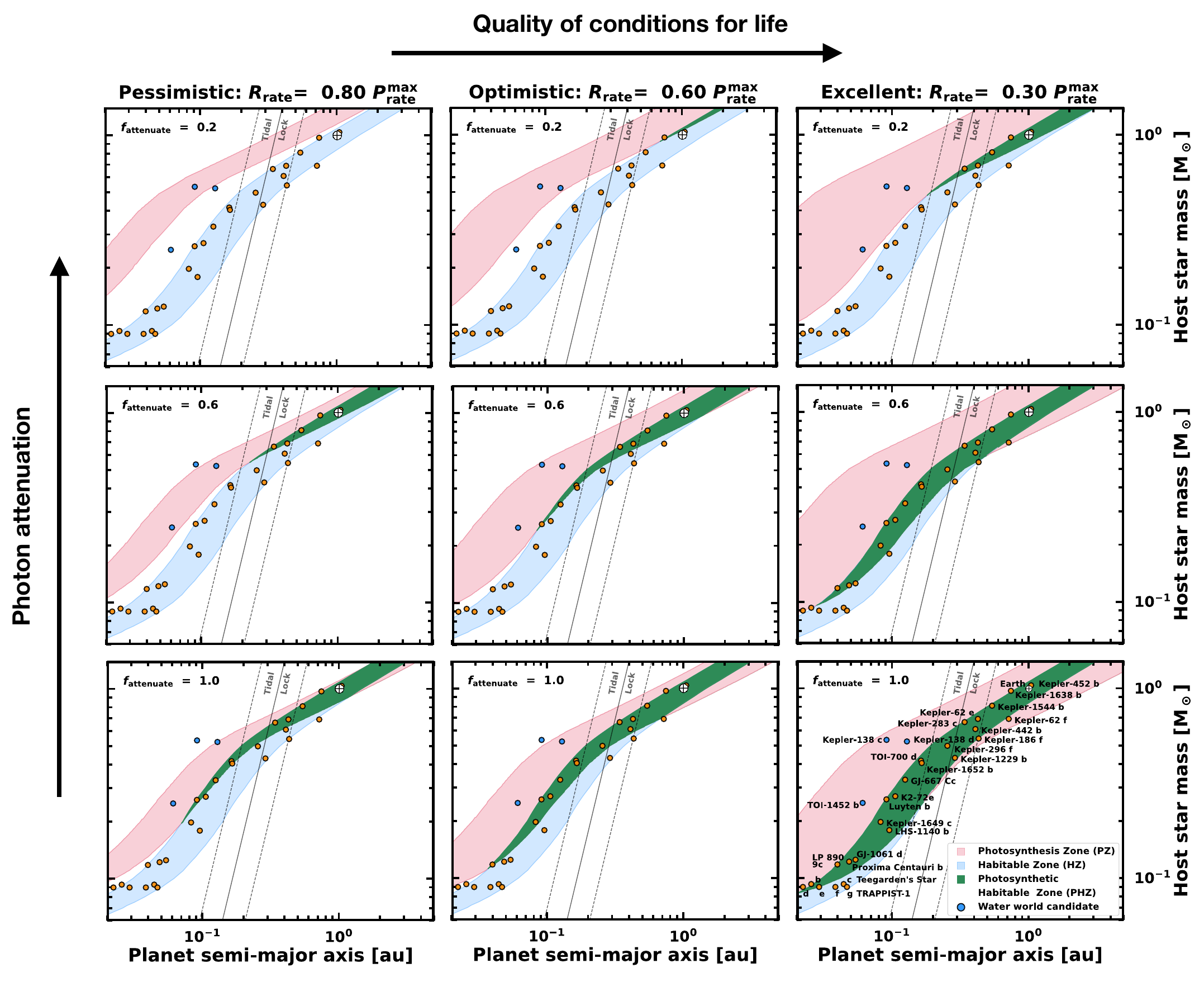}
    \caption{  \label{fig:bz}     The Photosynthetic Habitable Zone.  Region where positive net photosynthesis is possible shown in pink, and habitable zone shown in blue. The overlapping region is where photosynthesis can actually occur, and is named ``The photosynthetic habitable zone', since it is possible that oxygenic photosynthesis, and therefore biosignatures, could exist on planets in this location. Overplotted in yellow markers are planets of interest, i.e. planets in the habitable zone expected to have a solid surface ($R \lesssim 1.8$ R$_\mathrm{E}$). Blue markers indicate water world candidates. Earth is shown by white Earth symbol. The tidal lock radius with upper and lower limits is also shown.      }
\end{figure*}

We find the existence of the photosynthetic habitable zone (PHZ), shown in green in Figure \ref{fig:bz}), where photosynthesizing life could exist and therefore leave behind atmospheric biosignatures. It occurs where the $P(I,T)>0$ region (pink in Figure \ref{fig:bz}) overlaps with the region where liquid water can exist (the Habitable Zone, blue in Figure \ref{fig:bz}). 

Figure \ref{fig:bz} shows nine scenarios. On the $y$ axis of the whole plot, atmospheric attenuation of photons is increasing. On the $x$-axis of the whole plot, the quality of the conditions for life are increasing (i.e., the maximum photosynthetic rate is much higher than the baseline respiration rate). The ``Excellent'' column corresponds to conditions for photosynthesizing lifeforms on Earth with respiration rates at 30\% of maximum photosynthesis rates, typical for marine phytoplankton such as \textit{Isochrysis galbana},  \textit{Platymonas subcordiformis}, etc \citep{IPPOLITI201671,yang2020}.

 As the quality of conditions for life increases, the rate of respiration as a fraction of the  maximum photosynthetic rate attainable decreases, resulting in a larger PZ.  As atmospheric attenuation increases to Earth-like levels, interestingly, this decreases the size of the PZ, reducing the parameter space over which biosignatures could be found. This suggests that it may be easier for large-scale photosynthesizing organisms, such as cyanobacterial mats, to occur on planets with more tenuous atmospheres. Overplotted as yellow symbols are planets that spend at least 10\% of their orbit in the HZ and have radii $R_\mathrm{p}< 1.8$ R$_\mathrm{E}$, so could have a solid surface, since planets with $R_\mathrm{p}\gtrsim 1.8$ R$_\mathrm{E}$ are likely to be gas-dominated \citep{rocky1,rocky2}.

 Additionally, we also plot 3 recently identified water world candidates, Kepler-138 c and d \citep{piaulet2022} and TOI-1452 b \citep{cadieux2022}. While these fall outside the HZ, Kepler-138 d and TOI-1452 b fall inside the PZ. Although liquid water has not been directly detected on these planets, measurements imply that these planets may be similar to the water-rich icy moons of the solar system, such as Europa or Enceladus. These criteria reduce the $\sim$5000 known exoplanets down to 29 planets of interest.
 
 Overplotted on Figure \ref{fig:bz} is the tidal lock radius \citep{peale1977,kasting1993}:
\begin{equation}
r_{\mathrm{lock}}=0.027\left(\frac{P_{0} t}{Q}\right)^{1 / 6} M_*^{1 / 3},
\end{equation}
where $P_0$ is the original rotation period of the planet, $t$ is the age of the system (1 Gyr), $M_*$ is the stellar mass and $Q^{-1}$ is the solid body plus ocean specific dissipation function. We use $Q=100$, for the solid line, and $Q=10$ and $Q=1000$ for the upper and lower limits of the tidal lock radius. We assume $P_0=13.5$ hours, i.e. the day length of Earth when it was 1 Gyr old. 

There are several planets that are in or near the PHZ largely regardless of the quality of conditions for life or atmospheric effects, Kepler-452 b, Kepler-1638 b, Kepler-1544 b and Kepler-62 e and f. These planets should therefore be the most promising for detecting biosignatures.  It is also possible that some of these planets with $R\gtrsim 1.5$ R$_\mathrm{E}$ are water worlds \citep{luque2022}. The least promising candidates are those around the lowest mass host stars, $M_* \lesssim 0.4$  M$_\odot$, since their position in Figure \ref{fig:bz} coincides with the PHZ for less than half of our considered scenarios.


As a final result, we posit that meaningful discussions of the habitable zone should account for where photosynthesis is also possible, since almost all life on Earth depends on photosynthesis either directly or indirectly. We therefore suggest replacing the use of the phrase ``Habitable Zone'' with ``Photosynthetic Habitable Zone'', when searching for biosignatures.

\section{Discussion}
\label{sec:discussion}

\noindent Conventional photosynthesis, as experienced on Earth, takes place during the day via photosynthetically active radiation (PAR) received directly from the Sun. We do not consider here any other scenario, such as starlight PAR, moonlight PAR, planetlight PAR, or speculative biological adaptations. We work under the limiting assumption that life would exist approximately ``as we know it''. While other scenarios are possible in principle, they require increasingly complex caveats, such as an older universe or photosynthesis only occurring at full moon \citep[see, e.g.][]{ravencockell2006}.  Furthermore, we focus our consideration on atmospheric biosignatures, rather than biological surface features, since we expect atmospheric signatures to be detectable even when the disk-averaged spectrum features, such as the vegetation red edge \citep{seager2005}, are not \citep{cockell2009}.



Our analysis here intends to show a general trend rather than demarcate absolute boundaries, since the fit parameters ($\alpha,\beta$ and $\gamma$) in the $P-I$ relation can take a variety of values as long as the curve retains the same functional form, that is, an initial slope, a peak, and steady decline due to photoinhibition of photosynthesis at high intensity \citep[see, e.g.,][]{platt1976,platt1981,model3,ye2013}. 
%
Furthermore, atmospheric attenuation is a function of column density, which is a function of both planet mass and planet size. While a super Earth may have a more massive atmosphere than  Earth, a super Earth also has a larger surface area which results in atmospheric mass scaling more slowly than the increase in planet mass \citep{elkins-tanton2008}. Additionally, super Earth outgassing rates may be lower than the Earth's \citep{2012ApJ...748...41S}, so a more rigorous consideration of atmospheric attenuation must determine the role that planet mass plays.


It is unclear what effect tidal locking would have on the development of photosynthesizing life on another planet,  since much of life on earth depends on 24-hour circadian cycles to regulate physiological function \citep{dvornyk}. An absence of periodicity on tidally locked planets would likely impact the evolution of biological regulation in those systems.  The night side of the planet could not support photosynthesis since it does not receive PAR. This immediately discounts half of the surface area of the planet. On the other hand, always receiving PAR could potentially increase the rate of net O$_2$ production, since intensity does not wax and wane during the course of the day. In either case, it seems clear that there is a link between Earth's rotation rate and oxygenation, with longer days associated with higher oxygenation rates \citep{klatt2021}. Even if this is not the case, and the circadian rhythm for phototrophs simply originates in environmental behaviour rather than any advantage in photosynthetic production, our analysis indicates that the PHZ predominantly exists outside the tidal locking radius for all cases, suggesting that the search for life elsewhere in the Universe should be focused around non-tidally locked planets.




An important limitation of our work is that we do not consider the impact of temperature on photosynthesis rates. The planet equilibrium temperature depends on Bond albedo, which is a function of both stellar type \citep[e.g.][]{kopparapu2013}  and planet properties \citep{shields2013,rushby2019}, with variations up to an order of magnitude around an F type star depending on planet composition. Correctly calculating planet surface temperature therefore requires 1D energy-balance climate models, which will be explored in a future work (Hall et al. in prep). However, the HZ models that we use here have already placed bounds on surface temperature, such that there would be liquid water and therefore temperatures in the range $0-100\deg$ C.

Photosynthesis takes place on Earth in a variety of temperatures. In plants, it is generally constrained to lower temperature ranges ($10\degree$C - $40\degree$C) before suffering irreversible damage \citep{berry1980photosynthetic}, while in cyanobacteria the preferred range is somewhat higher, with the upper limit for non-thermophiles $\sim 73\degree$C \citep{ward2012ecology}.  A few points are worth noting - first, at low temperatures, photosynthesis is both enzyme-limited and phosphate limited due to a reduction in the availability of phosphate in chloroplasts, while at higher temperatures proteins become denatured. High or low temperatures could also affect the stability and fluidity of the cellular membranes in which photosynthesis machinery components localize, resulting in more or less efficient biochemical reactions at the membrane interface.

At moderate temperatures ($\sim$10-35$\degree$C), photosynthesis is mostly limited by the rate of CO$_2$ diffusion. This is another limitation of our work \textemdash cyanobacteria exist in mat-like colonies that have a $z$-depth, and we have assumed instead that any colony is essentially infinitesimally thin and we therefore do not need to consider diffusion equations. Another limitation of not considering the $z$-depth is that attenuation of photon flux by water occurs \citep{2021MNRAS.503.3434L}, affecting light availability to organisms found at different water column depths. Chloroplast-based photosynthesis in terrestrial plants largely depends on chlorophyll a for optimal absorbance of violet and orange light, while cyanobacteria possess additional light-harvesting phytochromes which allow light capture at wavelengths outside of optimal chlorophyll a absorbance \citep{kehoe2010}. If these phytoplankton exist solely underwater, the PHZ could therefore move closer to the central star, which may increase or decrease the size of the PHZ. Advanced modelling of microbial benthic ecology would be best suited to this problem, and we leave this to future work. 

We have also assumed that any extant life in the Universe shares a biochemistry similar enough to photosynthesizing lifeforms on Earth that we would recognise its signatures. Even on Earth, so-called ``exotic photosynthesis'' exists, such as infrared photosynthesis in anoxygenic photosynthetic organisms \citep{heath1999}. This anoxygenic photosynthesis uses hydrogen sulfide instead of water as the reductant, and produces sulphur instead of oxygen as a byproduct, e.g:
\begin{equation}
6 \mathrm{CO}_{2}+12 \mathrm{H}_{2} \mathrm{S}  \xrightarrow{h\nu}     \mathrm{C}_{6} \mathrm{H}_{12} \mathrm{O}_{6}+12 \mathrm{~S}+6 \mathrm{H}_{2} \mathrm{O}.
\end{equation}

The pigments used to carry out anaerobic photosynthesis are similar to chlorophyll, but have peak absorption in the near-IR due to molecular differences. While significant atmospheric sulphuric acid in this scenario could be considered a biosignature,  there is a large risk of false-positive results due to its occurrence in many non-biological processes \citep{domagal-goldman}, which is why it is not targeted as a biosignature. In addition to this, anoxygenic photosynthesis is likely an evolutionary precursor to oxygenic photosynthesis, with biogeochemical changes on a terrestrial planet forcing a switch to an oxygen producing version \citep{raven2007,raven2009functional,raven2009contributions}.
%

One thing that we do not consider here is the effect of planet mass on atmospheric composition and density. For example, a super-Earth that is outside the HZ, that retains its primordial H-He dominated atmosphere could have surface temperatures that are warm enough to host liquid water \citep{mollous2022}. The same could therefore also be true of regions that we have determined too cold for net positive photosynthesis. On the other hand, we also determine in this work having less atmospheric attenuation results in a broader PHZ, which counteracts the positive effect of the retained H-He atmosphere. 


A further interesting avenue of exploration is regarding the impact of CO$_2$ on planetary temperature through the greenhouse effect, along with the impact on photosynthetic rates. An inherent assumption of our work is that photosynthesis is not limited by reduced CO$_2$ availability, because the fluctuations on Earth are small. The global average concentration of CO$_2$ today is 400 ppm, and was significantly higher when life first emerged due to Earth's secondary atmosphere.  Space-based observatories show the CO$_2$ concentration varies only at $\sim$few ppm levels \citep{2016GeoRL..4311400H,2019RemS...11..850H}, whereas rate limitation requires a decrease of $\sim$tens of ppm below this 400ppm level \citep[e.g.][]{moss1962,gabrielsen1948}. However, the flip side of this is that we have not explored what this means for atmospheres rich in CO$_2$, which would be likely to increase the expected surface temperature as a function of instellation. A self-consistent CO$_2$-instellation habitable zone should be calculated for this.

 Finally, it could be possible that most habitable worlds in the Universe simply have no detectable signs of life \citep{cockell2014},  either because they are uninhabited, are too young to have evolved life yet ($<1$ Gyr based on the Earth's fossil record), have biotic chemistry at concentrations too low to detect, or the biotic atmospheric chemistry is indistinguishable from the abiotic. 
 
\subsection{Planet rotation periods}
Cyanobacteria appeared in the fossil record $\sim$3.5 Gyr ago,  just $\sim$1 Gyr after the Earth's formation. The concentration of O$_2$ remained at primordial values of $\lesssim$10$^{-3}$ of present atmospheric levels (PALs) until $\sim$2.1 Gyr, when a dramatic increase in atmospheric O$_2$ occurred \textemdash the so-called Great Oxidation Event (GOE). Earth's rotation period is currently 24 hours having been slowed by tidal interaction with the moon, but is likely to have been as low as 6 hours 4 Gyr ago \citep{lambeck1980earths,cuk2012making,bartlett2016analysis}. Earth's rotation period may therefore have increased by more than a factor of two since the evolution of photosynthesis. 

Recently, it has been postulated that the GOE occurred when Earth's daylength increased to $\sim$16 hours \citep{klatt2021}. To test this hypothesis, \citet{klatt2021} performed numerical simulations of movement of O$_2$ in cyanobacterial mats for daylengths between 12 and 52 hours,  using simulated diel light cycle illumination. They found that longer days resulted in higher net O$_2$ flux through the cyanobacteria mat, and verified these results by taking measurements from real cyanobacterial colonies in controlled conditions. This lead them to conclude that  increases in daylength could plausibly have influenced Earth’s oxygenation, particularly around key oxidation events, and thus helped to pave the way for the evolution of plants and animals as we know them. 

At 1 Gyr, Earth's daylength was $\sim$ 13 hours with  gravitational modelling \citep{moon} predicting a spin-down to 16 hour days by $\sim$1.9 Gyr (late Archean). Within the next 0.3 Gyr the atmospheric O$_2$ concentration increased to $\sim$0.1 PAL.  Assuming Earth-like  biology, the work of \citet{klatt2021} suggests that a planetary period of $\gtrsim$16 hours could be an important factor in producing an oxygen-rich atmosphere to support life as we know it.

\begin{table*}[!ht]
    \centering
    \begin{tabular}{lllllllll}
    \hline
        
        Planet & a [au] &  M$_*$ [M$_\odot$] & M [M$_{\mathrm{E}}$] & M$_\mathrm{est.}$ [M$_{\mathrm{E}}$]  &   R [R$_{\mathrm{E}}$] & P[days] & day [hrs] & day [hrs]\\
        
           & &  &  &  ($R = M^{0.27}$) &    & & Calculated& Estimated\\
           \hline
           \hline
        

        GJ-1061 d & 0.054 & 0.125 & 1.64 & 1.73 & 1.16 & 13.00 & TL & TL \\  
        GJ-667 Cc & 0.125 & 0.330 & 3.81 & 4.95 & 1.54 & 28.10 & TL & TL \\  
        K2-72e & 0.106 & 0.270 & 2.21 & 2.57 & 1.29 & 24.20 & TL & TL \\  
        Kepler-1229 b & 0.290 & 0.430 & - & 3.48 & 1.40 & 86.80 & - & 11.39 \\  
        Kepler-138 c & 0.091 & 0.535 & 2.3 & 4.60 & 1.51 & 13.78 & TL & TL \\  
        Kepler-138 d & 0.129 & 0.525 & 2.1 & 2.03 & 1.21 & 23.09 & TL & TL \\  
        Kepler-1544 b & 0.542 & 0.810 & - & 8.46 & 1.78 & 168.80 & - & 9.28 \\  
        Kepler-1638 b & 0.745 & 0.970 & - & 10.16 & 1.87 & 259.30 & - & 8.90 \\  
        Kepler-1649 c & 0.083 & 0.198 & - & 1.24 & 1.06 & 19.50 & - & 14.43 \\  
        Kepler-1652 b & 0.165 & 0.404 & - & 5.70 & 1.60 & 38.10 & - & 10.16 \\  
        Kepler-186 f & 0.432 & 0.544 & - & 1.79 & 1.17 & 129.90 & - & 13.27 \\  
        Kepler-283 c & 0.341 & 0.664 & - & 9.19 & 1.82 & 92.70 & - & 9.11 \\  
        Kepler-296 f & 0.255 & 0.498 & - & 8.82 & 1.80 & 63.30 & - & 9.19 \\  
        Kepler-442 b & 0.409 & 0.610 & - & 3.04 & 1.35 & 112.30 & - & 11.75 \\  
        Kepler-452 b & 1.048 & 1.037 & 5 & 6.11 & 1.63 & 384.80 & 11.057 & 10.00 \\  
        Kepler-62 e & 0.427 & 0.690 & - & 5.83 & 1.61 & 122.40 & - & 10.11 \\  
        Kepler-62 f & 0.718 & 0.690 & - & 3.57 & 1.41 & 267.30 & - & 11.32 \\  
        LHS-1140 b & 0.096 & 0.179 & 6.38 & 6.25 & 1.64 & 24.70 & TL & TL \\  
        LP 890-9 c & 0.040 & 0.118 & - & 3.21 & 1.37 & 8.46 & - & 11.60 \\  
        Luyten b & 0.091 & 0.260 & 2.89 & 8.82 & 1.80 & 18.65 & TL & TL \\  
        Proxima Centauri b & 0.049 & 0.122 & 1.27 & 2.64 & 1.30 & 11.19 & TL & TL \\  
        Teegarden's b & 0.026 & 0.093 & 1.05 & 1.08 & 1.02 & 4.91 & TL & TL \\  
        Teegarden's c & 0.044 & 0.093 & 1.11 & 1.16 & 1.04 & 11.40 & TL & TL \\  
        TOI-1452 b & 0.061 & 0.249 & 4.82 & 6.71 & 1.67 & 11.06 & TL & TL \\  
        TOI-700 d & 0.163 & 0.416 & 1.72 & 1.62 & 1.14 & 37.40 & TL & TL \\  
        TRAPPIST-1 d & 0.022 & 0.090 & 0.39 & 0.40 & 0.78 & 4.05 & TL & TL \\  
        TRAPPIST-1 e & 0.029 & 0.090 & 0.69 & 0.73 & 0.92 & 6.10 & TL & TL \\  
        TRAPPIST-1 f & 0.038 & 0.090 & 1.04 & 1.16 & 1.04 & 9.21 & TL & TL \\  
        TRAPPIST-1 g & 0.047 & 0.090 & 1.32 & 1.57 & 1.13 & 12.40 & TL & TL \\
        \hline
    \end{tabular}
         \caption{\label{tab:planets}Properties of candidate exoplanets in the photosynthetic habitable zone assuming excellent conditions. Daylengths are estimated using the empirical relation of Equation~\ref{eq:period}, which is not valid if the planet is tidally-locked (TL). If the mass is not known, it is estimated using Eq. \ref{eq:mass-radius}. Values from the NASA exoplanet archive.}
\end{table*}

Unfortunately, the rotation period, or daylength, is measured only for a handful of exoplanets (2M1207 b, PSO J318.5, GQ Lup b and $\beta$ Pic b), all of which are massive, fast rotators \citep[for a summary see][]{scholz2018}. An empirical spin-mass relationship, where equatorial velocity is given  by $v_\mathrm{eq}\propto\sqrt{M}$, is known to fit the solar system planets. However, both Mercury and Venus do not fit this trend due to tidal interactions with the Sun, and the Earth's tidal interactions with the moon also result in deviation from this relationship.
The solar system trend is:

 
\begin{equation}
v_\mathrm{eq} = A ({M}/{\mathrm{M_J}})^{\frac{1}{2}},
\end{equation}
with $v_\mathrm{eq}$ in units of km s$^{-1}$ and $A=13.1$. The rotation period of the planet in seconds is then
\begin{equation}
T = \frac{2 \pi R}{A} \bigg( \frac{M}{\mathrm{M_J}}\bigg)^\frac{1}{2}
\end{equation}
where $R$ is the planet radius. If the exoplanet radius is known (or estimated), then the period of the planet in 24-hour days is

\begin{equation}
\label{eq:period}
T = 0.632\left(\frac{M_{\mathrm{E}}}{M}\right)^{\frac{1}{2}}\left(\frac{R}{R_{\mathrm{E}}}\right), 
\end{equation}
and can therefore be predicted directly. 
\begin{figure}
    \centering
    \includegraphics[width=\linewidth]{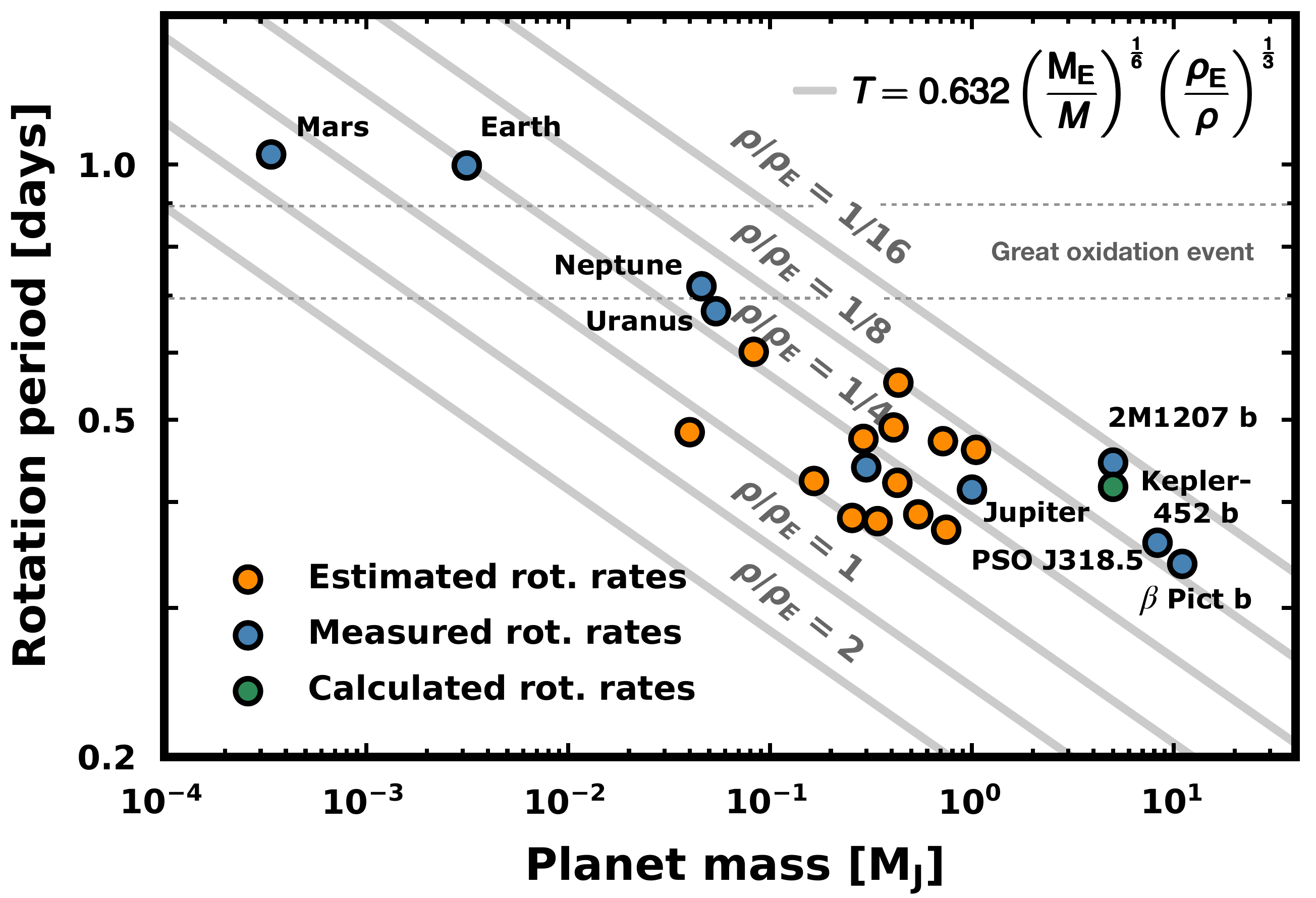}
    \caption{  \label{fig:period}     Mass-period relationship. Day lengths of planets either measured directly (blue points) or estimated through empirically determined relationship in Eq. \ref{eq:period} (green and yellow points). Yellow points indicate that mass was also estimated using Eq. \ref{eq:mass-radius}. Grey solid lines are plots of Equation \ref{eq:period-radiusunknown} for a plausible range of planet densities. Horizontal dashed lines mark the range of Earth's daylength predicted for the GOE \citep{klatt2021}.}
   %
\end{figure} 

However, for small radii planets (as we consider here), they generally do not also have associated mass measurements, but estimates can be made using a mass-radius relationship. We take the mass and radius values for low mass planets with an Earth-like composition from Table 1 of \citet{fortney2007},  which gives us a best-fit mass-radius relation of
\begin{equation}
\label{eq:mass-radius}
R_\mathrm{p} = M_\mathrm{p}^{0.27}
\end{equation}
with a high coefficient of determination $R^2 = 0.98$. We use this to estimate planet mass, given in the fifth column of Table \ref{tab:planets}. For Kepler-452 b, we calculate the rotation rate using the observed mass and retrieve 11 hours. For all other planets in our sample, we use the mass estimate from Eq. \ref{eq:mass-radius}. The relationship is shown in Figure \ref{fig:period}. If longer daylengths are required for atmospheric oxygenation, then our sample of planets may not yet be rotating slowly enough. If, however, their rotation has been slowed due to moons, then these planets may have the right conditions for their own GOE.


It is useful to note that alternatively, if the radius is unknown and the mass is known, then a density can be assumed, and we can instead write:
\begin{equation}
\label{eq:period-radiusunknown}
T=0.632\left(\frac{M_{\mathrm{E}}}{M}\right)^{\frac{1}{6}}\left(\frac{\rho_{\mathrm{E}}}{\rho}\right)^{\frac{1}{3}},
\end{equation}

A tidally-locked planet will experience constant daylight on one half of its surface, and constant darkness (or reflected moonlight \citealt{lingam2019}) on the other, and will therefore not experience the diurnal variation in light intensity of non tidally-locked planets. It is unclear whether this could be helpful or harmful to oxygenic photosynthesis. On the one hand, constant illumination means photosynthesis is always possible as long all other conditions allow, and on the other hand, dark respiration is not. A key question is therefore: is there an advantage to the light-dark cycle for life? \citet{Tang2000}  explored this in an experiment probing the effects of daylight length and temperature on arctic cyanobacteria. The total daylength was held constant at 24 hours, and they varied the length of daytime, $L$, between 8 and 24 hours for three fixed temperatures of 5, 15, and 25$^\circ$C. The cyanobacteria growth rates increased with increasing L for 5$^\circ$C, but they plateaued with $L$ for 10$^\circ$C and 15$^\circ$C, resulting in a reduction in net photosynthesis that was largest for the 24 hour daylight case. Unfortunately, the experimental errors on the measured respiration rates were too large to discriminate between the different conditions. 

However, other work has shown that dark respiration rates are at their peak shortly after the transition from lightness to darkness, and steeply decline as the period of darkness increases \citep{markager1992}. In a similar vein, peak O$_2$ production (rather than respiration) was found to occur at different times in the 24 hour and 52 hour daylengths of \citet{klatt2021}. Peak O$_2$ production occurred before noon in the 52 hour daylength and after noon in the 24 hour daylength. More work is therefore required to determine how tidal locking may impact the photosynthesis-respiration cycle for the tidally locked planets in this work. 

\section{Conclusion}
\label{sec:conclusion}

We have demonstrated the existence of a photosynthetic habitable zone (PHZ). It is the distance from the host star where the habitable zone overlaps with where photosynthesis is possible. We argue that the search for biosignatures of oxygenic photosynthesizing life forms should be concentrated in the PHZ if we expect photosynthesis in the Universe to proceed in a similar manner to photosynthesis on Earth. The PHZ becomes smaller with increasing atmospheric attenuation (i.e., more dense atmospheres), and so may make life less likely on super-Earths, since their larger gravitational field can hold onto more atmosphere. The PHZ also becomes smaller as the conditions for life become less favourable, which we describe as respiration rate relative to maximum possible photosynthetic rate, increasing. We therefore conclude that the parameter space for signs of life is far narrower than the standard HZ.

Out of the nine scenarios we considered, we found TRAPPIST 1-e to be in the photosynthetic habitable zone for one scenario - little atmospheric affects and excellent conditions for life. However, it is almost certainly tidally locked, and it is not clear how or if photosynthetic life can proceed on tidally locked planets. Furthermore, the global circulation models of tidally-locked planets by \citet{lobo22} find that the HZ is limited to a narrow strip along the terminator for water-limited rocky planets. This reduces both the fraction of planet surface area for liquid water and cyanobacteria mats, and potentially the amount of water for photosynthesis. It may therefore not be the best place to focus the search for signs of life. We identify five planets, Kepler-452 b, Kepler-1638 b, Kepler-1544 b and Kepler-62 e and Kepler-62 f, that are consistently in the PHZ in a variety of environments. For Kepler-452 b, we calculate that it should have a rotation period of 11 hours. The other four planets are estimated to have rotation periods between 9 and 11 hours. We suggest the search for signs of life elsewhere in the Universe should begin in earnest on the candidate planets we have identified.


\section{Acknowledgements}
With special thanks from CH to Duncan H. Forgan \citep{forgan2019solving}. We thank the referee for their work reviewing this manuscript. CH also thanks Ken Rice, Nancy Kiang and Ethan Van Woerkom for insightful discussion. This study was supported in part by resources and technical expertise from the Georgia Advanced Computing Resource Center, a partnership between the University of Georgia's Office of the Vice President for Research and Office of the Vice President for Information Technology. This work has made use of the NASA Exoplanet Catalogue {\url{https://exoplanets.nasa.gov/discovery/exoplanet-catalog/}, and the exoplanet archive \url{https://exoplanetarchive.ipac.caltech.edu/}

\bibliographystyle{aasjournal}
\bibliography{bib}{}

\label{lastpage}
\end{document}